\documentclass[prl,twocolumn,superscriptaddress]{revtex4}

\usepackage{graphicx}
\usepackage{amssymb}
\usepackage{times}
\usepackage{amsmath}

\begin{document}

\title{Quantum State Engineering with 
Continuous-Variable Post-Selection
}

\author{Andrew~M.~Lance} \affiliation{Quantum Optics Group, Department
of Physics, Faculty of Science, Australian National University,
ACT 0200, Australia}

\author{Hyunseok~Jeong} \affiliation{Department of Physics, University of Queensland, St Lucia,
Queensland 4072, Australia}

\author{Nicolai~B.~Grosse} \affiliation{Quantum Optics Group, Department
of Physics, Faculty of Science, Australian National University,
ACT 0200, Australia}
\author{Thomas~Symul} \affiliation{Quantum Optics Group, Department
of Physics, Faculty of Science, Australian National University,
ACT 0200, Australia}

\author{Timothy~C.~Ralph} \affiliation{Department of Physics, University of Queensland, St Lucia, 
Queensland 4072, Australia}

\author{Ping~Koy~Lam} \affiliation{Quantum Optics Group, Department
of Physics, Faculty of Science, Australian National University,
ACT 0200, Australia}

\date{\today}

\begin{abstract}

We present a scheme to conditionally engineer an optical 
quantum system via continuous-variable measurements. 
This scheme yields 
high-fidelity squeezed single photon and superposition 
of coherent states, from input single and two photon Fock states
respectively. The input Fock state
is interacted with an ancilla squeezed vacuum state
using a beam splitter. We transform the quantum system
by post-selecting on the continuous-observable 
measurement outcome of the ancilla state. 
We experimentally demonstrate 
the principles of this scheme using 
coherent states and measure experimentally 
fidelities that are only achievable using 
quantum resources.

\end{abstract}


\maketitle

{\it Introduction - }
The transformation or engineering of quantum states 
via measurement induced conditional evolution is an 
important technique for discrete variable systems, 
particularly in the field of quantum information~\cite{NIE00}
Typically, the quantum system of interest 
is interacted with a prepared ancilla state, which
is then measured in a particular basis. 
The system state is retained, or discarded,
depending on the measurement outcome,
resulting in the controlled conditional 
evolution of the quantum system.
It is a necessary condition for inducing 
a non-trivial conditional evolution that the 
interaction of the ancilla and system produces 
entanglement between them.
  
In optical systems, highly non-linear evolutions, 
which are difficult to induce directly, can be induced conditionally 
on systems 
by post-selecting on particular photon 
counting outcomes~\cite{OBR03}. 
In principle, a near deterministic, 
universal set of unitary transformations 
can be induced on optical qubits in this way~\cite{KNI01}. 
Importantly, it was shown that arbitrary optical states can 
be engineered conditionally, based on discrete 
single photon measurements~\cite{DAK99}. 

Recently, there has been increased interest in conditional 
evolution based on continuous-variable 
measurements~\cite{Lau03, Bab05, RAL05}.  
In these schemes a quantum system is interacted 
with a prepared ancilla, which is measured via 
a {\it continuous} observable, e.g. the amplitude or 
phase quadratures of the electromagnetic field.  
This has been experimentally demonstrated for 
a system using a beamsplitter as the interaction 
and a vacuum state as the ancilla, with conditioning 
based on homodyne detection~\cite{Bab05}. 
A similar system using conditioning 
of adaptive phase measurements has also been 
studied~\cite{RAL05}. 

In this letter, we investigate a continuous-variable 
conditioning scheme based on a beam splitter interaction, 
homodyne detection and an ancilla squeezed vacuum state. 
We theoretically show that for input one and two photon
Fock states, this scheme yields high fidelity 
squeezed single photon Fock states and 
superposition of coherent states (SCS) 
respectively,  which are highly non-classical 
and interesting quantum states that have potentially 
useful applications in quantum information 
processing~\cite{catapply}. 
We experimentally demonstrate the principles of this 
scheme using input coherent states, and 
measure experimental fidelities that
are only achievable using quantum resources. 

\begin{figure}[htbp]
\begin{center}
\includegraphics[width=\columnwidth]{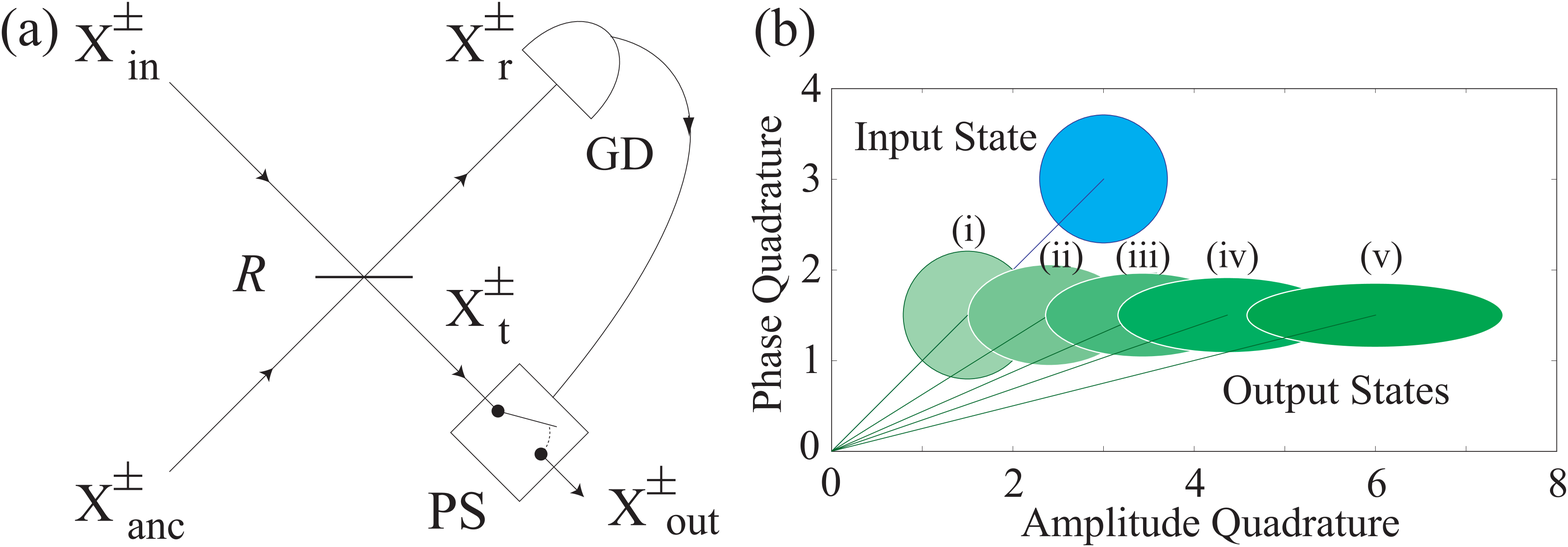}
\caption{(Color online) (a) Schematic of the post-selection protocol. 
$\hat{X}_{\rm in}^{\pm}$: amplitude (+) and phase (-) 
quadratures of the input state; (anc) 
ancilla state; (r) reflected; (t) transmitted;  and (out)
post-selected output state. 
$R$: beam splitter reflectivity; GD: gate detector;
PS: post-selection protocol.
(b) Standard deviation contours of the 
Wigner functions of an 
input coherent state (blue) and post-selected 
output states (green) for 
$R=0.75$ and varying ancilla state 
squeezing of (i) $s=0$,
(ii) $s=0.35$, (iii) $s=0.69$, 
(iv) $s=1.03$ and (v) ideal squeezing. 
}\label{ExptSetup}
\end{center}
\end{figure}

{\it Theory - }
The squeezed vacuum ancilla state used in our scheme
is represented as ${\hat S}(s)|0\rangle$
with the squeezing operator ${\hat S}(s) = {\rm exp}[-(s/2)(\hat{a}^2 - \hat{a}^{\dagger 2})]$,
where $s$ is the squeezing parameter and $\hat a$ is the annihilation operator.
The Wigner function of the squeezed vacuum is 
$W_{\rm sqz}(\alpha;s)=2\exp[-2(\alpha^{+})^2e^{-2s}-2(\alpha^{-})^2e^{2s}]/\pi$,
where $\alpha=\alpha^{+}+i\alpha^{-}$ with real quadrature variables
$\alpha^{+}$ and $\alpha^{-}$.
The first step of our transformation protocol is to interfere 
the input field with the ancilla state
on a beam splitter as shown in Fig.~\ref{ExptSetup}~(a).
The beam splitter operator ${\hat{B}}$ acting on
 modes $a$ and $b$ is represented as 
$\hat{B}(\theta)=\exp \{(\theta /2)
(\hat{a}^{\dagger }\hat{b}
-\hat{b}^{\dagger }\hat{a})\}$, 
where the reflectivity is defined as 
$R=\sin^2(\theta /2)$ and $T=1-R$.
A homodyne measurement is performed on the amplitude
quadrature on the reflected field mode, with the measurement
result denoted as $X^+_{\rm r}$. 
The transmitted state is post-selected for $|X^+_{\rm r}|<x_0$, where
the post-selection threshold $x_0$ is 
determined by the required fidelity
between the output state and the ideal target state.

We first consider a single-photon state input, $|1\rangle$,
and a squeezed single photon, 
${\hat S}(s^\prime)|1\rangle$, as the target state.
The Wigner function of the single photon state 
is
$W^{|1\rangle}_{\rm in}(\alpha)=2\exp[-2|\alpha|^2](4|\alpha|^2-1)/\pi$.
After interference via 
the beam splitter, the resulting two-mode state becomes 
$W(\alpha,\beta)=W^{|1\rangle}_{\rm in}\big( \sqrt{T}\alpha+\sqrt{R}\beta\big)
W_{\rm anc}\big(-\sqrt{R}\alpha+\sqrt{T}\beta\big)$,
where $W_{\rm anc}= W_{\rm sqz}(\alpha;s)$ and
$\beta=\beta^{+}+i\beta^{-}$.
The transmitted state after the homodyne detection of 
the reflected state is
$W_{\rm out}(\alpha;X^+_{\rm r})=P_1(X^+_{\rm r})^{-1}\int_{-\infty}^{\infty}
d\beta^{-}W(\alpha,\beta^{+}=X^+_{\rm r},\beta^{-})$, where the 
normalization parameter is 
$P_1(X^+_{\rm r})=\int_{-\infty}^\infty d^2\alpha
d\beta^{-}W(\alpha,\beta^{+}=X^+_{\rm r},\beta^{-})$.
If the measurement result is $X^+_{\rm r}=0$, 
the Wigner function of the output state becomes 
\begin{eqnarray}\label{eq:Wout1}
W_{\rm out}(\alpha)
&=&\frac{2}{\pi}e^{-2[e^{-2s^\prime}(\alpha^{+})^2+e^{2s^\prime}(\alpha^{-})^2]}\\ \nonumber
&&\times\big(4e^{-2s^\prime}(\alpha^{+})^2+4e^{2s^\prime}(\alpha^{-})^2-1\big), 
\end{eqnarray}
where
$s^\prime=-\ln [(T + e^{-2s}R)^2]/4$.
One can immediately notice that the output 
state in Eq.~(\ref{eq:Wout1})
is {\it exactly} the Wigner function of a 
squeezed single photon, $\hat{S}(s^\prime)|1\rangle$.
We note that the output squeezing $s^\prime$ 
can be arbitrarily close to 
the squeezing of the ancilla state $s$ by making $R$ close to zero.
For the nonzero post-selection threshold criteria 
$|X^+_{\rm r}|<x_0$, the corresponding success probability
is given by $P_s(x_0)=\int_{-x_0}^{x_0} d X^+_{\rm r} P_1(X^+_{\rm r})$. 

\begin{figure}[htbp]
\includegraphics[width=\columnwidth]{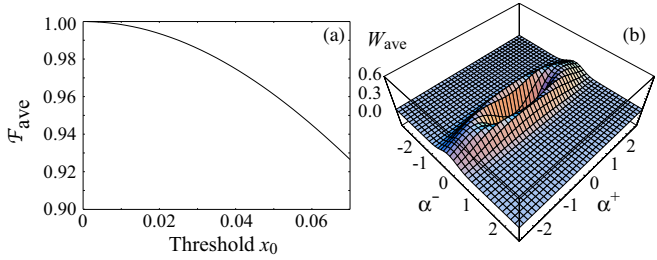}
\caption{
(Color online) (a) The average fidelity ${\cal F}_{\rm ave}$
between the post-selected output state of an 
input single photon state $|1\rangle$,
and the squeezed single photon state 
${\hat S}(s^\prime)|1\rangle$, 
for varying threshold $x_0$.   
The beam splitter reflectivity is $R=0.98$,
ancilla state squeezing is $s=0.7$
and target state squeezing
is $s^\prime=0.67$.
(b) The average Wigner function 
$W_{\rm ave}$ of the output state for 
$F_{\rm ave}=0.99$, 
for $x_0=0.025$ and $P_s=0.003$.
}
\label{1photon_Input}
\end{figure}

We calculate the fidelity between the output state (with 
the measurement result $X^+_{\rm r}$)
and the ideal target state given by 
${\cal F}_1(X^+_{\rm r})=\pi\int^\infty_{-\infty}
d^2\alpha W_{\rm out}(\alpha;X^+_{\rm r})W_{\rm out}(\alpha)$. 
From this we can determine the 
average fidelity for the threshold $x_0$ defined as 
${\cal F}_{\rm ave}(x_0)=\int_{-x_0}^{x_0} dx P_1(X^+_{\rm r}){\cal F}_1(X^+_{\rm r})/
\int_{-x_0}^{x_0}d{\tilde X^+_{\rm r}}
P_1({\tilde X^+_{\rm r}})$. We use this 
average fidelity measure to characterize the efficacy 
of our protocol for nonzero thresholds. 
We can likewise determine the 
average Wigner function $W_{\rm ave}(\alpha;x_0)$
for the threshold $x_0$.

Fig.~\ref{1photon_Input}~(a) shows the average fidelity 
${\cal F}_{\rm ave}$ for varying post-selection
threshold $x_0$. This figure illustrates that high fidelity 
squeezed single photon states can be produced 
using experimentally realizable squeezing of the 
ancilla state and finite thresholds. The average Wigner function
corresponding to an average fidelity  ${\cal F}_{\rm ave}=0.99$
is shown in Fig.~\ref{1photon_Input}~(b).
We point out that post-selection around $X^{+}_{\rm r}=0$
preserves the non-Gaussian features of the input state.
Our scheme enables one to perform 
the squeezing of a single photon with high fidelities 
using any {\it finite} degree of squeezing of the ancilla state. 
We point out that this squeezed single photon state is a 
good approximation to an odd SCS, which has applications 
in quantum information processing \cite{catapply}. 
We emphasize that this interesting result
{\it cannot} be achieved by continuous 
electro-optic feed-forward methods 
~\cite{Fil05} with finite ancilla state squeezing. 

\begin{figure}[htbp]
\includegraphics[width=\columnwidth]{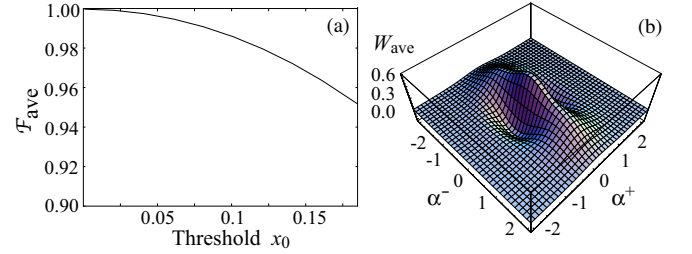}
\caption{
(Color online) (a) The average fidelity ${\cal F}_{\rm ave}$
between the output state 
of the input two photon state, $|2\rangle$,
and the ideal superposition of coherent 
states (SCS), for varying threshold $x_0$.
The beam splitter reflectivity is $R=1/2$, the 
ancilla state squeezing is $s=-0.37$, and
the amplitude of the SCS is $\gamma=1.1i$.
(b) The average Wigner function 
$W_{\rm ave}$ of the output state
for $F_{\rm ave}=0.99$, for 
$x_0=0.084$ and $P_s=0.052$.}
\label{2photon_Input}
\end{figure}

Another example of our 
post-selection protocol is for the case of 
input  two-photon Fock states, $|2\rangle$.
In this case, our target state is an even SCS,
$|\gamma\rangle+|-\gamma\rangle$ (unnormalized),
where $|\gamma\rangle$ is a coherent state of amplitude 
$\gamma=\gamma^{+}+i\gamma^{-}$.
The Wigner representation of the SCS is
\begin{eqnarray}
W_{\rm scs}(\alpha)=N_{1}\Big\{e^{-2|\alpha-\gamma|^2}
+e^{-2|\alpha+\gamma|^2}~~~~~~~~~~~~~~~~~~~~~~~~\nonumber\\
+ e^{-2|\gamma|^2}(e^{-2(\alpha+\gamma)^*(\alpha-\gamma)}
+e^{-2(\alpha+\gamma)(\alpha-\gamma)^*})
\Big\},
\end{eqnarray}
where 
$N_{1}=\{\pi(1+ e^{-2|\gamma|^2})\}^{-1}$.
For an input two photon Fock state, 
the fidelity between the post-selected 
output state (with 
the measurement result $X^+_{\rm r}$)
and the ideal SCS target state is
${\cal F}_2(X^+_{\rm r})=\pi\int^\infty_{-\infty}
d^2\alpha W_{\rm out}(\alpha;X^+_{\rm r})W_{\rm scs}(\alpha)$.
From this expression, the average fidelity 
$\mathcal{F}_{\rm ave}$ and average Wigner 
function $W_{\rm ave}$ for a post-selection
threshold $x_0$ can be be calculated. 
Fig.~\ref{2photon_Input} shows the average fidelity 
of the output state for varying threshold, which
illustrates that high fidelity SCS 
can be obtained with experimentally 
realizable ancilla state squeezing and finite thresholds.
The average Wigner function
corresponding to an average fidelity of ${\cal F}_{\rm ave}=0.99$
is shown in Fig.~\ref{2photon_Input}~(b). 
Once such SCSs are obtained, they can be 
conditionally amplified for SCSs of larger amplitudes
using only linear optics  schemes~\cite{Lund04}.

We now consider the case of a Gaussian state, i.e.,
an {\it unknown} coherent state, $|\gamma\rangle$, as the input. 
The post-selection scheme for $X^+_{\rm r}=0$ transforms the coherent state as 
\begin{equation}
D(\gamma)|0\rangle\longrightarrow 
D\big(\sqrt{T}[e^{2s^\prime}
\gamma^{+}+i \gamma^{-} ]\big)S(s^\prime)|0\rangle,
\label{eq:ppt}
\end{equation}
where $D(\gamma)={\rm exp}[\gamma\hat{a}^\dagger-\gamma^{*}\hat{a}]$ 
is the displacement operator.
Figure~\ref {ExptSetup}~(b) illustrates that the squeezing and 
the displacement transformation of the post-selected output state
 in Eq.~(\ref{eq:ppt}) is dependent 
on the ancilla state squeezing, and that the 
output state is a minimum uncertainty state independent 
of the ancilla state squeezing. In the limit of ideal 
ancilla state squeezing, 
the post-selection scheme works as an ideal 
single-mode squeezer for arbitrary input coherent states
$D(\gamma)|0\rangle\rightarrow S(s^\prime)D(\gamma)|0\rangle $.
In this case, the output squeezing is 
$s^\prime\rightarrow-\ln [T]/2$.
We note that for input coherent states this scheme provides
an alternative method of squeezing to electro-optic protocols 
presented in~\cite{Fil05}.

\begin{figure}[htbp]
\begin{center}
\includegraphics[width=\columnwidth]{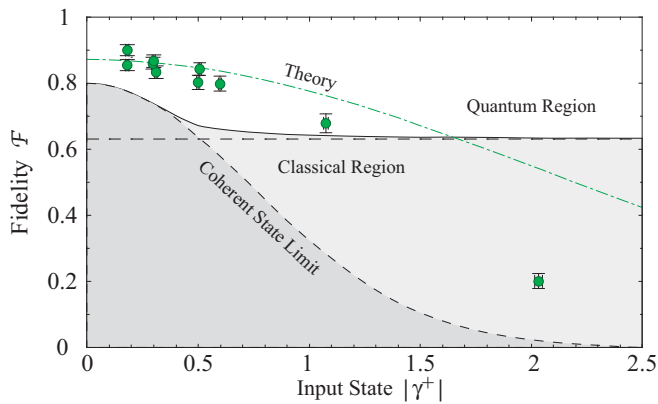}
\caption{
(Color online) Experimental fidelity for varying 
amplitudes $|\gamma^{+}|$ of the input
coherent state, for $R=0.75$ and
$x_0= 0.01$ Dark grey region: classical fidelity limit for an
ancilla vacuum state; Light grey region: classical fidelity limit.
Dot-dashed line: calculated theoretical prediction of experiment, 
with experimental losses and inefficiencies.
}\label{ExptFidelity}
\end{center}
\end{figure}

{\it Experiment - }
We experimentally demonstrated the principle of the 
post-selection protocol using input displaced coherent
states for a realizable ancilla state squeezing.
For the experiment, the quantum states we considered
reside at the sideband frequency $(\omega)$ of the 
electromagnetic field. We denote the quadratures 
of these quantum states as 
$\hat{X}^{\pm}=\langle\hat{X}^{\pm}\rangle+\delta\hat{X}^{\pm}$, 
where $\langle\hat{X}^{\pm}\rangle$ are the mean quadrature 
displacements, and where the quadrature variances are expressed 
by $V^{\pm}=\langle(\delta\hat{X}^{\pm})^{2}\rangle$. 
 
Fig.~\ref{ExptSetup} shows the experimental setup. 
We used a hemilithic MgO:LiNbO$_3$ 
below-threshold optical parametric amplifier, to
produce an amplitude squeezed field at 1064~nm
with squeezing of  $s=0.52\pm0.03$, corresponding to
a quadrature variance of $V^{+}_{\rm anc}=-4.5\pm 0.2$~dB 
with respect to the 
quantum noise limit. More detail of this experimental production 
of squeezing is given in~\cite{Bow03}. 
The displaced coherent states were produced at the 
sideband frequency of 6.81~MHz of a laser field at 1064~nm, using
standard electro-optic modulation techniques~\cite{Bow03}. 
The post-selection protocol goes as follows: 
the amplitude squeezed ancilla field $\hat{X}^{\pm}_{\rm anc}$
was converted to a phase squeezed field 
by interfering it with the input coherent 
state $\hat{X}^{\pm}_{\rm in}$ with a much larger 
coherent amplitude on the beam splitter 
with a relative optical phase shift of $\pi/2$. 
This optical interference 
yielded two output states that were phase squeezed. 
The optical fringe visibility between the two fields was 
$\eta_{\rm vis}=0.96\pm 0.01$. 

We directly detected the amplitude quadrature of the 
reflected state $\hat{X}^{+}_{\rm r}$ 
using a gate-detector, which
had a quantum efficiency of $\eta_{\rm det}=0.92$ and 
an electronic noise of $6.5$~dB below the 
quantum noise limit. The post-selection could, in principle, be 
achieved using an all optical setup, but here
we post-selected {\it a posteriori} the  
quadrature measurements of the transmitted state, 
$\hat{X}^{\pm}_{\rm t}$, 
which were measured using a balanced homodyne 
detector. The total homodyne detector efficiency was
$\eta_{\rm hom}=0.89$, with the electronic noise
of each detector $8.5$~dB below the quantum noise limit.
To characterize the protocol, we also measured 
the quadratures of the 
input coherent state, $\hat{X}^{\pm}_{\rm in}$, 
using the same homodyne detector. 
To ensure accurate results, the total homodyne detector 
inefficiency was inferred out 
of all quadrature measurements~\cite{Bow03}. 

The electronic photocurrents of the detected quantum 
states (at a sideband frequency of 6.81~MHz) from the gate 
and homodyne detectors were electronically 
filtered, amplified and demodulated down to 25~kHz
using an electronic local oscillator at 6.785~MHz. 
The resulting photocurrents were 
digitally recorded using a NI PXI 5112 data acquisition system at a sample 
rate of 100~kS/s. 
We used computational methods to filter, 
demodulate and down-sample the data, so 
that it could be directly analyzed 
in the temporal domain. 
From this data, we post-selected the 
quadrature measurements 
of the transmitted state, $\hat{X}^{\pm}_{\rm t}$, 
which satisfied the threshold criteria 
$|X^{+}_{\rm r}|<x_0$. 
This post-selection threshold was 
independent of the input state and was
experimentally optimized depending on the 
beam splitter reflectivity. 

We characterized the efficacy of our 
 protocol as an ideal single mode squeezer, by
determining the fidelity of the post-selected output state 
with a target state that is an ideal squeezed operation 
of the input state [Eq.~(\ref{eq:ppt})]. The Wigner function of this 
ideal squeezed input state is given by 
$W_{\rm out}(\gamma;s^\prime)$, where 
$s\rightarrow\infty$ and $s^\prime\rightarrow-\ln [T]/2$. 
In this case, the fidelity is given by ${\cal F}(X^+_{\rm r})=
\pi\int^\infty_{-\infty} d^2\gamma W_{\rm expt}(\gamma;X^+_{\rm r})
W_{\rm out}(\gamma;s^\prime)$, where 
$W_{\rm expt}(\gamma;X^+_{\rm r})$ is the 
Wigner function of the post-selected output state. 
From this expression the average fidelity 
${\cal F}_{\rm ave}$ for a post-selection threshold 
$x_0$ can be calculated. 
This corresponds to unity fidelity 
$\mathcal{F}_{\rm ave}=1$ in the limit of ideal ancilla state 
squeezing and $X^+_{\rm r}=0$.
In the experiment,  the input state was 
a slightly mixed state due to inherent 
low-frequency classical noise on the laser beam, 
with quadrature variances of $V^{+}_{\rm in}=1.13\pm 0.02$ 
and $V^{-}_{\rm in}=1.05\pm 0.02$, with respect to the 
quantum noise limit. Hence, we calculated
the fidelity of the post-selected output state
with an ideal squeezed transform 
of the experimental input state. 
Fig.~\ref{ExptFidelity} shows the 
classical fidelity limit 
$\mathcal{F}_{\rm clas}$, which signifies the highest 
fidelity achievable when the interaction 
of the ancilla and input coherent state yields 
no entanglement. Exceeding this classical fidelity 
limit can only be achieved using quantum resources.
 
Fig.~\ref{ExptFidelity} shows the experimental fidelity 
for varying input states 
$|\gamma^{+}| \equiv |\langle\hat{X}^+_{\rm in}\rangle|$.
For a beam splitter reflectivity of $R=0.75$, 
we achieved a best fidelity of $\mathcal{F}_{\rm ave}=0.90\pm0.02$ for
an input state $|\gamma^{+}|=0.18\pm0.01$, which 
exceeds the maximum classical fidelity of $\mathcal{F}_{\rm clas}=4/5=0.8$.
This post-selected output state had
quadrature variances of $V^{+}_{\rm out}=4.70\pm0.11$ 
and $V^{-}_{\rm out}=0.51\pm0.01$. The mean quadrature 
displacement gains, $g^{\pm}=\langle\hat{X}^{\pm}_{\rm out}\rangle/
\langle\hat{X}^{\pm}_{\rm in}\rangle$, were measured to be
$g^{+}=0.71\pm0.16$ and  $g^{-}=0.50\pm0.06$. This is
compared with the ideal case of perfect ancilla 
state squeezing, where the ideal theoretical gains are 
$g^{+}_{\rm ideal}=2$ and $g^{-}_{\rm ideal}=1/2$. 
The phase gain was controlled by 
the beam splitter transmittivity, whilst 
the amplitude gain was less the ideal case  
due to finite ancilla state squeezing, 
finite post-selection threshold
and experimental losses. 

\begin{figure}[htbp]
\begin{center}
\includegraphics[width=\columnwidth]{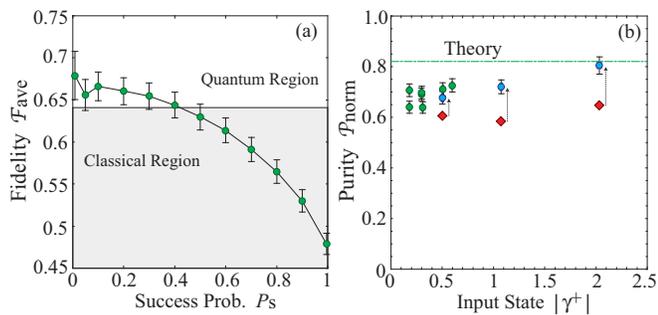}
\caption{(Color online) (a) Experimental fidelity ($\mathcal{F}_{\rm ave}$) for varying 
post-selection success probability, for 
an amplitude $|\gamma^{+}|=1.07$ of the input state and 
$R=0.75$.  
(b) Experimental purity ($\mathcal{P}_{\rm norm}$) 
for varying input state
$|\gamma^{+}|$, for $x_0= 0.01$ 
Dash arrows show purity 
prior to (diamonds) and 
after (circles) post-selection. 
Dot-dashed line: calculated theoretical 
prediction of the experiment.}\label{ExptFidelityPurity}
\end{center}
\end{figure}

The quantum nature of the post-selection 
protocol is demonstrated by the experimental fidelity 
results that exceed the classical fidelity limit in 
Fig.~\ref{ExptFidelity}.
For large input states $|\gamma^{+}|$, 
the experimental fidelity was less than the
theoretical prediction due to
electronic detector noise and the finite 
resolution of the data acquisition 
system, resulting in a smaller 
post-selected output state 
$|\gamma^{+}|$
and a corresponding 
decrease in the experimental fidelity.
Fig.~\ref{ExptFidelityPurity}~(a) illustrates how
the experimental fidelity 
of a post-selected state transitions to the 
quantum fidelity region by 
decreasing the post-selection
threshold (and corresponding 
probability of success). 

We also characterized the experiment in terms of 
the purity of the post-selected output state, defined 
as $\mathcal{P}={\rm tr}(\rho^{2}_{\rm out})$. In the case of 
Gaussian states, the purity of the output state can be expressed as
$\mathcal{P}=(V^{+}_{\rm out}V^{-}_{\rm out})^{-1/2}$. 
In the ideal case of a lossless experiment and a 
post-selection threshold $X^+_{\rm r}=0$, the  protocol is a 
purity preserving transform, independent of the input state 
and the amount of squeezing of the ancilla state. 
In the experiment, as the input states are slightly mixed, 
we calculate the purity of the post-selected output state, 
normalized to the purity of the input state, which is given by 
$\mathcal{P}_{\rm norm}=
(V^{+}_{\rm out}V^{-}_{\rm out})^{-1/2}/(V^{+}_{\rm in}V^{-}_{\rm in})^{-1/2}$.
Fig.~\ref{ExptFidelityPurity}~(b) shows the experimental purity 
of the post-selected output state for varying input states, 
which illustrates how the purity is improved via the 
post-selection process.
For a beam splitter reflectivity of $R=0.75$ we achieved a best 
purity of $\mathcal{P}_{\rm norm}=0.81\pm0.04$ 
for an input state of $|\gamma^{+}|=2.03\pm0.02$.
Fig.~\ref{ExptFidelityPurity}~(b) shows that  
the purity of the post-selected output states 
were approximately independent of the input states,
for a large range of input states.

We also implemented our scheme  
for a beam splitter reflectivity of $R=0.5$. In this case,
we measured a best fidelity of $\mathcal{F}_{\rm ave}=0.96\pm0.01$, 
which exceeded the maximum classical fidelity of 
$\mathcal{F}_{\rm clas}=\sqrt{8}/3\approx0.94$, and
measured a best purity of 
$\mathcal{P}_{\rm norm}=0.80\pm0.04$. 

In summary, we have investigated a continuous-variable 
conditioning scheme based on a beam splitter interaction, 
homodyne detection and an ancilla squeezed vacuum state. 
The conditional evolution of quantum systems based on 
continuous-variable measurements of the ancilla state 
are of particular interest as they can yield from input Fock states, 
non-Gaussian states, which have applications
in the field of quantum information. Further, for 
Gaussian states, this technique 
provides an alternative to continuous electro-optic 
feed-forward schemes.
We theoretically showed that our
conditional post-selection scheme yields high fidelity
squeezed single photon and superposition of 
coherent states from input one and two photon
Fock states respectively, for realizable squeezing of the ancilla state.
We experimentally demonstrated the principles of this 
scheme using coherent states, and 
measured experimental fidelities that
were only achievable using quantum resources.

We thank the Australian Research Council for financial 
support through the Discovery Program. 

\end{document}